\documentclass[prd,twocolumn,nofootinbib,preprintnumbers,floatfix]{revtex4}
\usepackage[utf8]{inputenc}
\usepackage{amsfonts,amsmath,amssymb}
\usepackage{graphicx}
\usepackage{color}
\usepackage[plainpages=false, colorlinks=true, anchorcolor=blue, linkcolor=blue, citecolor=blue, bookmarks=false]{hyperref}
\usepackage{natbib}
\usepackage{enumitem}
\usepackage{subcaption}
\newcommand{\rthis}[1]{\textcolor{black}{#1}}
\begin{document}
\newcommand{\apjl}{Astrophys. J. Lett.}
\newcommand{\apjs}{Astrophys. J. Suppl. Ser.}
\newcommand{\aap}{Astron. \& Astrophys.}
\newcommand{\aj}{Astron. J.}
\newcommand{\araa}{Ann. Rev. Astron. Astrophys. } %ARA$\&$A}
\newcommand{\aapr}{Astronomy and Astrophysics Review}
\newcommand{\mnras}{Mon. Not. R. Astron. Soc.}
\newcommand{\apss} {Astrophys. and Space Science}
\newcommand{\jcap}{JCAP}
\newcommand{\pasj}{PASJ}
\newcommand{\LRR}{Living Reviews in Relativity}
\newcommand{\pasa}{Pub. Astro. Soc. Aust.}
\newcommand{\pasp}{PASP}
\newcommand{\physrep}{Physics Reports}
\newcommand{\ssr}{Space Science Reviews}

\title{A test of constancy of dark matter halo surface density and radial acceleration relation in relaxed galaxy groups }
\author{Gopika \surname{K.}}\altaffiliation{E-mail:ph19resch01001@iith.ac.in}

\author{Shantanu  \surname{Desai}}  
\altaffiliation{E-mail: shntn05@gmail.com}

\begin{abstract}
The dark matter halo surface density, given by the product of the dark matter core radius ($r_c$) and core density ($\rho_c$) has been shown to be a constant for a wide range of isolated galaxy systems. 
Here, we carry out a test of this {\em ansatz} using a sample of 17 relaxed galaxy groups observed using Chandra and XMM-Newton, as an extension of  our previous analysis with galaxy clusters. We find that  $\rho_c \propto r_c^{-1.35^{+0.16}_{-0.17}}$,  with an intrinsic scatter of about 27.3\%, which is about 1.5 times larger than that seen for galaxy clusters. Our results thereby indicate that the surface density is discrepant with respect to scale invariance by about 2$\sigma$, \rthis{and its value is about four times greater than that for galaxies. Therefore,  the elevated values of the halo surface  density for groups and clusters indicate  that the surface density cannot be a universal constant for all dark matter dominated systems.}
Furthermore, we  also implement a test of the radial acceleration relation  for this group sample.   We find that the residual scatter in the radial acceleration  relation is about 0.32 dex and a factor of three larger than that obtained using galaxy clusters.  The acceleration scale  which we obtain is  in-between that seen for  galaxies and clusters.

\end{abstract}

\affiliation{Department of Physics, Indian Institute of Technology, Hyderabad, Telangana-502285, India}
\maketitle

\section{Introduction}
In a recent work~\cite{Gopika_2020} (GD20, hereafter), we  carried out an observational test using a sample of 12 relaxed galaxy clusters observed using the Chandra X-ray observatory~\cite{Vikhlinin06}, to ascertain  if the dark matter halo surface density is constant.  This was motivated by the \rthis{claim from some groups} that  the dark matter halo surface density has been observed to be  invariant,  for a wide variety of systems spanning over 18 orders in blue magnitude,  for a diverse suite of systems from spiral galaxies to dwarf galaxies~\cite{Donato,Kormendy14,Burkert15,Salucci19}. 
 This dark halo surface density ($S_{surf}$) is given by the product of core radius ($r_c$) and core density($\rho_c$)\footnote{Note that $\rho_c$ is also sometimes referred to as central density~\cite{Kormendy14}}.
 \begin{equation}
 S_{surf} = \rho_c \times r_c 
 \label{eq:S}
 \end{equation}
 Both $r_c$ and $\rho_c$
were obtained by fitting a Burkert~\cite{Burkert95} or other  cored profiles, or in a model-independent fashion as done in GD20 or~\cite{Chan}. The current best-fit values for the dark matter  surface density  for single galaxy systems  is given by $\log  (\rho_c r_c)=(2.15 \pm 0.2)$ in units of  $M_{\odot} pc^{-2}$~\cite{Salucci19}.

\rthis{The main premise behind the above results is that all dark matter profiles are cored. However, the  cored profiles do not provide a pristine fit to all types of systems from dwarf galaxies to clusters, some of which need cuspy profiles~\cite{Simon,Strigari,Genina,Helmi,Newman,Newman2}}. 
Therefore, this result has  been disputed by other authors,  who  argue that the halo surface density is correlated with the luminosity, mass, and other galaxy properties ~\cite{Boyarsky,Napolitano,Cardone, DelPopolo12,Cardonedel,Saburova,DelPopolo20}.

\rthis{Given these conflicting results}, a further test of the constancy of dark matter halo surface density for a large suite of astrophysical systems would be an acid test for  $\Lambda$CDM   as well as various alternatives. 
For example, it  has  been shown recently  that the observed constant surface density is in tension with predictions from fuzzy dark matter models, thus ruling them out~\cite{Burkert20}.
A recent recap of the predictions of some of the myriad  theoretical scenarios for the halo surface density can be found in GD20. Prior to GD20, there was only one work~\cite{Chan}, which implemented this test for a large sample of galaxy clusters and showed that the power-law exponent for the relation between $\rho_c$ and $r_c$ ranges from -1.46 to -1.6, instead of -1, which is expected for a constant halo density.  GD20 then showed using the Chandra cluster dataset that $\rho_c \propto r_c^{-1.08 \pm 0.55}$, indicating that the halo surface density is almost invariant. GD20  showed that they could reproduce the earlier result~\cite{Chan}  for the Chandra dataset by  neglecting the stellar  and gas mass contribution to the total mass, similar to the analysis in ~\cite{Chan}. However, the observed dark matter halo surface density in GD20 (and also~\cite{Chan})  is about an order of magnitude higher ($\log(r_c \rho_c) \sim 3$ in units of  $M_{\odot} pc^{-2}$) than what has been deduced for  single galaxy systems~\cite{Salucci19}. \rthis{Therefore, at face value this clearly shows the dark matter surface density of haloes cannot be universal (constant) at all scales,  as claimed in ~\cite{Donato}.} 

It is still an open question, \rthis{on whether the elevated halo surface density for galaxy clusters} is consistent with hydrodynamical simulations of $\Lambda$CDM and alternatives ~\cite{Gopika_2020}. Therefore, given the observed increase in surface density for cluster scale haloes compared to galaxies, it is imperative  to carry out  this test for astrophysical systems  in an intermediate mass range, which bridge the gap between single galaxies and clusters. Groups provide the ideal laboratory for this purpose.

Another intriguing observational result was found by  ~\cite{Mcgaugh16,Lelli}, who obtained a pristine deterministic relation between the baryonic ($a_{bar}$) and total acceleration ($a_{tot}$) from spiral galaxies in the SPARC galaxy sample,  with the scatter attributed to observational uncertainties. A similar relation has also been observed for elliptical galaxies~\cite{Sheth}.  This has been dubbed as  Radial Acceleration Relation (RAR).
\begin{equation}
a_{tot} = \frac{a_{bar}}{1-e^{-\sqrt{a_{bar}/a_0}}},
\label{eq:RAR}
\end{equation}
where $a_0 \sim 1.2 \times 10^{-10} m/s^2$~\cite{Mcgaugh16}. More generally, this relation can be recast as a linear regression between $a_{tot}$ and $a_{bar}$ in log-log space~\cite{Tian,Chan20}.
A few groups have disputed the existence of this acceleration scale using the same data or found higher scatter using other samples~\cite{delpopolo18,Rodrigues20,courteau,Chang}. Nevertheless, this relation can be shown to be  a trivial consequence of the MOND paradigm~\cite{Famaey12}. There are conflicting results in literature on whether  $\Lambda$CDM simulations and semi-analytical models can reproduce the acceleration scale and the tight scatter (See ~\cite{Paranjpe} for an up-to-date recap of most of these predictions). Most recently,~\cite{Paranjpe} has shown that they can reproduce the observed scatter in the RAR from  quasi-adiabatic relaxation of dark matter haloes in the presence of baryons.
Although, it is known for more than three decades that the MOND phenomenology cannot obviate the need for dark matter in galaxy clusters~\cite{The88,Sowmya}, a test of RAR would be an acid test for the models which reproduce MOND behavior at galactic scales but need dark matter at cluster scales. Motivated by these considerations, multiple groups~\cite{Chan20,Tian,Pradyumna,Pradyumna2} carried out a test of the RAR using different samples of galaxy clusters. These  analyses found a tight scatter of between 0.11 to 0.13 dex and an elevated acceleration scale $a_0 \sim 10^{-9} m/s^2$. This acceleration scale is again about an order of magnitude larger than that seen for galaxies in the SPARC sample. 

As a follow-up to these sets of tests, we now carry out a  test of both a constant dark matter halo surface density and RAR for galaxy groups. Galaxy groups are less massive, gravitationally  bound systems compared to galaxy clusters, with intra-group temperatures less than 2 keV~\cite{Sun12,Eke}. However, they cannot be considered as  scaled down versions of massive clusters, only because of their shallow gravitational potential. Groups are much more susceptible to complex baryonic physics.  For these reasons,  groups have proved to be excellent laboratories for understanding the baryonic physics  needed for understanding galaxy formation~\cite{Fino,Sun09,Sun12,Reiprich,Morandi}.

Although, MOND has been tested (and disfavored) with X-ray selected relaxed groups~\cite{Angus_2008}, there has been no test of RAR or constant halo surface density for galaxy groups. Since  galaxy clusters point to an acceleration scale and  halo surface density,  about an order of magnitude higher than for galaxies~\cite{Gopika_2020,Pradyumna}, galaxy groups provide the ideal dataset to bridge   the gap between the two.   Combining the surface density data of groups with cluster enables us to increase the dynamic range to test the scaling with $M_{200}$, as predicted in some theoretical models~\cite{DelPopolo12,Loeb}.

This manuscript is structured as follows. We describe
the G07 group sample and associated models for the mass and acceleration profile in Section~\ref{sec:data}. Our analysis and results for the relation between core radius and density  and the radial acceleration relation (RAR) can be found in Section~\ref{sec:analysis} and Section~\ref{sec:RAR}, respectively. We also test for a dependence of  the dark matter scale parameters with mass and luminosity in Section~\ref{sec:corr}.  \rthis{Our results on the dark matter column density  (another variant of the surface density) are presented in Section~\ref{sec:sburk}.}
We conclude in Section~\ref{sec:conclusions}.

% \begin{equation}
% \rho_c r_c = 41 M_{\odot} pc^{-2} \times \left(\frac{M_{200}}{10^{10}M_{\odot}}\right)^{0.18}
% \label{loebeq}
% \end{equation}
% Del Popolo et al also predicted~\cite{DelPopolo12,DelPopolo17} a similar relation between the dark matter column density and $M_{200}$, within the context of a secondary infall model~\cite{Delpopolo09} valid for cluster scale haloes
% \begin{equation}
% \log (S)= 0.16  \log \left(\frac{M_{200}}{10^{12}M_{\odot}}\right)+2.23
% \label{delpopoloeq}
% \end{equation}

\section{Data Sample}
\label{sec:data}
We use   X-ray observations of 17 galaxy groups with redshifts upto $z=0.08$  from \citeauthor{Gastaldello_2007} (G07) and \citeauthor{Zappacosta_2006} (Z06) for this work. These groups were imaged with Chandra and/or XMM-Newton. The Chandra satellite has a high spatial resolution, which helps in resolving the temperature and density profiles in the cores. On the other hand, the high sensitivity of XMM-Newton provides very good signal-to-noise (S/N) ratios in the outer regions. 
Their exposure times ranged from 10-75 ks.
These groups  with masses ($10^{13}-10^{14}M_\odot$) span the range between galaxies and clusters, with  temperatures between 1-3 keV.  The groups were imaged upto outer radii of 730 kpc (cf. Table 1 in G07). More details of the observations and data reduction can be found in the aforementioned works~\cite{Gastaldello_2007,Zappacosta_2006}.
The dataset in G07 and Z06  consists of the  brightest and the most relaxed groups, which were selected in order to derive robust constraints on the mass profiles.
This  allows us to apply the equation of  hydrostatic equilibrium
to obtain the masses. The only exception is RGH 80, which has been proposed to be a sub-merging group~\citep{mahdavi2005xmm}. In G07, it has been included as it is part of a complete X-ray flux limited sample observed by Chandra and so that it can be used for comparing against the very relaxed systems. 
%The Chandra and XMM-Newton observation data were compiled by \citeauthor{Gastaldello_2007} and the XMM-Newton observation was used by \citeauthor{Zappacosta_2006} for the X-ray analysis. 
The X-ray brightness and temperature was fitted with 3-D parametric models described in G07. More detailed description about these parametric gas density and temperature models for this sample of galaxy groups can be found  in G07 and Z06.
\\
\subsection{Mass and Density in Groups}
\label{sec:procedure}
As the groups we choose for this analysis are relaxed systems, we can apply the equation of  hydrostatic equilibrium in these systems in order to determine their total gravitating mass. Errors due to  hydrostatic equilibrium assumption could  contribute a  systematic error of upto $\sim 15$\%~\cite{Biffi}. 
Hence, the total mass of the galaxy groups can be derived, given the temperature and gas density models and also assuming an ideal gas equation of state as~\cite{Mantz}
\begin{equation}
M(r)=-\frac{kT(r)r}{G\mu m_p}\bigg(\frac{d \ln \rho_g}{d \ln r} + \frac{d \ln T}{d \ln r}\bigg)
\label{eq:eq1}
\end{equation}
The 3-D temperature and gas density data were  obtained by fitting the X-ray data with the parametric models available in G07. These temperature and gas density data thus obtained were used to derive the mass using Eq.~\eqref{eq:eq1}. The temperature and density points at various radii for each of these groups were made available to us (F. Gastaldello, private communication).
We used  a spline interpolation to determine the derivatives of the logarithm of  the gas density and temperature needed to evaluate Eq.~\ref{eq:eq1}. The only exception was  A2589, for which  we used the best-fit values of the parametric models available in Z06, to derive the temperature and gas density profiles, and thereby the mass.
%The temperature and gas density are obtained by fitting the X-ray data with the parametric models described in G07\citep{Gastaldello_2007}.

Once we have the total mass, we then subtract the baryonic mass  in order to obtain the dark matter mass: 
\begin{eqnarray}
 M_{DM}=M(r)-M_{baryon}
 \\ \text{where}, \quad M_{baryon}=M_{gas}+M_{BCG}
 \label{eq:eq2}
\end{eqnarray}
Here, $M_{gas}$ and $M_{BCG}$ denote the mass of the gas and Brightest Cluster galaxy (BCG) respectively. The gas mass can be obtained from  the gas density assuming spherical symmetry:
\begin{equation}
    M_{gas}=\int 4\pi r^2 \rho_g(r)dr
    \label{eq:eq3}
\end{equation}
For most of the groups in this sample, the stellar contribution from the non-central galaxies is below 10\% except for AWM 4 and A262. In order to account for the stellar counterpart, we subtract  a BCG constant mass  from the total mass, similar to the analysis in \cite{Angus_2008}. The mass of the BCG  is determined with the assumption that $M/L_K =1$, where $L_K$  is the luminosity in the $K$-band.
This  {\it ansatz} is  valid  for the older stellar populations and a Kroupa IMF~(\citep{Kroupa2001}). However, for a Salpeter IMF, $M/L_k$ ratio can increase upto 50\%~\cite{Humphrey}. Note however that in G07,  the stellar mass has been modeled using a De Vaucouleurs profile. The $K$-band luminosities for all the groups  have been obtained from G07 and \cite{Angus_2008}.

\par The density of the dark matter halos can be obtained from the mass distribution  by positing spherical symmetry:
 \begin{equation}
\rho_{DM} (r) =\frac{1}{4\pi r^2}\frac{dM_{DM}}{dr}
\label{eq:eq4} 
\end{equation}
The errors in  the temperature and gas density for the selected radii for the groups in G07 were also  made available to us,(F. Gastadello, private correspondence) which  were used to propagate the errors in mass and density of the dark matter halos. For A2589, the errors were determined by propagating the errors of the best-fit parameters of the temperature and gas density models in Z06. The errors in the spherical symmetry assumption could be up to 5\%, as discussed in GD20 and references therein. These were not included in our error budget. We also do not consider the errors due to the uncertainty in the stellar mass distribution in our analysis.

The dark matter core density ($\rho_c$) and core radius ($r_c$) is obtained by fitting the halo density to a Burkert profile~\cite{Burkert95,Donato},
 \begin{equation}
\rho(r) =\frac{\rho_c r_c^{3}}{(r^2+r_c^2)(r+r_c)}
\label{eq:eq5} 
\end{equation}
%Note that unlike GD20,  we do not use the model independent technique for estimating $\rho_c$ and $r_c$, since we do not have many data points close to the group center for a robust extrapolation to zero radius.

\subsection{Acceleration Profile in Groups}
In order to test the  RAR for groups, we need to determine the total acceleration as well as  that  due to only the baryonic content at different radii from the center. The total acceleration was synthesized from the total mass (obtained from Eq.~\ref{eq:eq1}),
 \begin{equation}
a_{tot}=\frac{G M(r)}{r^2}
\label{eq:eq6} 
\end{equation}
Similarly, it is straight-forward to determine the acceleration due to the baryons alone, given the baryonic mass (obtained from Eq.~\ref{eq:eq2}). Thus, we have,
 \begin{equation}
a_{baryon}=\frac{G M_{baryon}}{r^2}
\label{eq:eq7}
\end{equation}

Again, we  assume spherical symmetry for the calculation of total mass. The acceleration was determined at the same radii for which  the temperature and density values were available. These radii are not the same for all the  groups used in our analysis. We used a  total  of 178 data points for all the groups. 
%SD: mention how many data points were used
\section{Analysis and Results}
\label{sec:analysis}
We now discuss our results on the constancy of  the halo surface density and RAR.
\subsection{Determination of scaling relations}
\begin{figure}
    \includegraphics[scale=0.50]{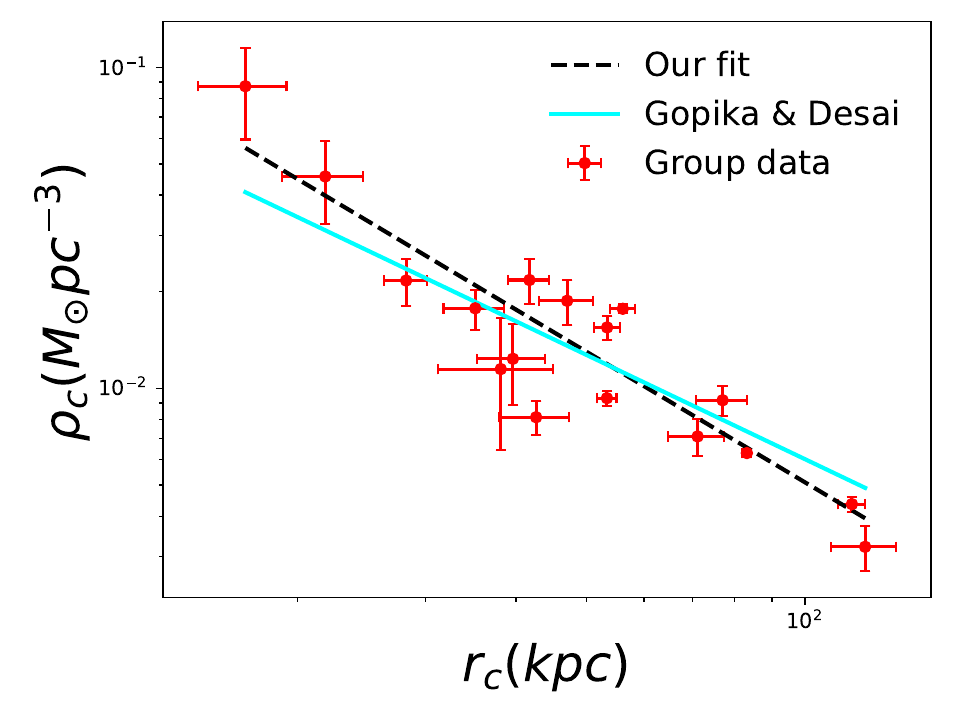}
    \caption{$\rho_c$ versus $r_c$ for the Z06 and G07 group sample. The black dashed line represents the fitted line from this analysis ($\rho_c \propto r_c^{-1.35}$). The cyan line shows the fit obtained for the clusters analyzed in GD20~\cite{Gopika_2020}.}
    \label{fig:f1}
\end{figure}

The $\rho_c$ and $r_c$ values along with the $1\sigma$ errors for the sample of 17 groups determined by fitting the density profiles to Eq.~\ref{eq:eq5}, as prescribed in Section~\ref{sec:procedure} are tabulated in Table \ref{tab:table1}.
The core density and core radius are plotted on log-log  scale  in Figure \ref{fig:f1}.  We find (similar to previous works) that the core density is inversely proportional to the core radius. We now determine the scaling relation between $\rho_c$ and $r_c$ in the same way as in GD20,  by fitting for $y=mx+b$, where $y \equiv \ln (\rho_c)$ and $x \equiv \ln (r_c)$.

We estimated the best-fit parameters of $m$ and $b$ by maximizing the logarithm of likelihood function given by:
\begin{eqnarray}
-2\ln L &=& \large{\sum_i} \ln 2\pi\sigma_i^2 + \large{\sum_i} \frac{[y_i-(mx_i+b)]^2}{\sigma_i^2}
\label{eq:eq8}  \\
\sigma_i^2 &=& \sigma_{y_i}^2+m^2\sigma_{x_i}^2+\sigma_{int}^2\nonumber
\end{eqnarray}

An unknown intrinsic scatter ($\sigma_{int}$), considered as a free parameter, is added in quadrature to the errors in $x$ ($\sigma_x$) and $y$ ($\sigma_y$), similar to  our previous works\cite{Gopika_2020,Pradyumna,Bora_feb}.

\begin{figure*}
    %\centering
    \includegraphics[scale=0.85]{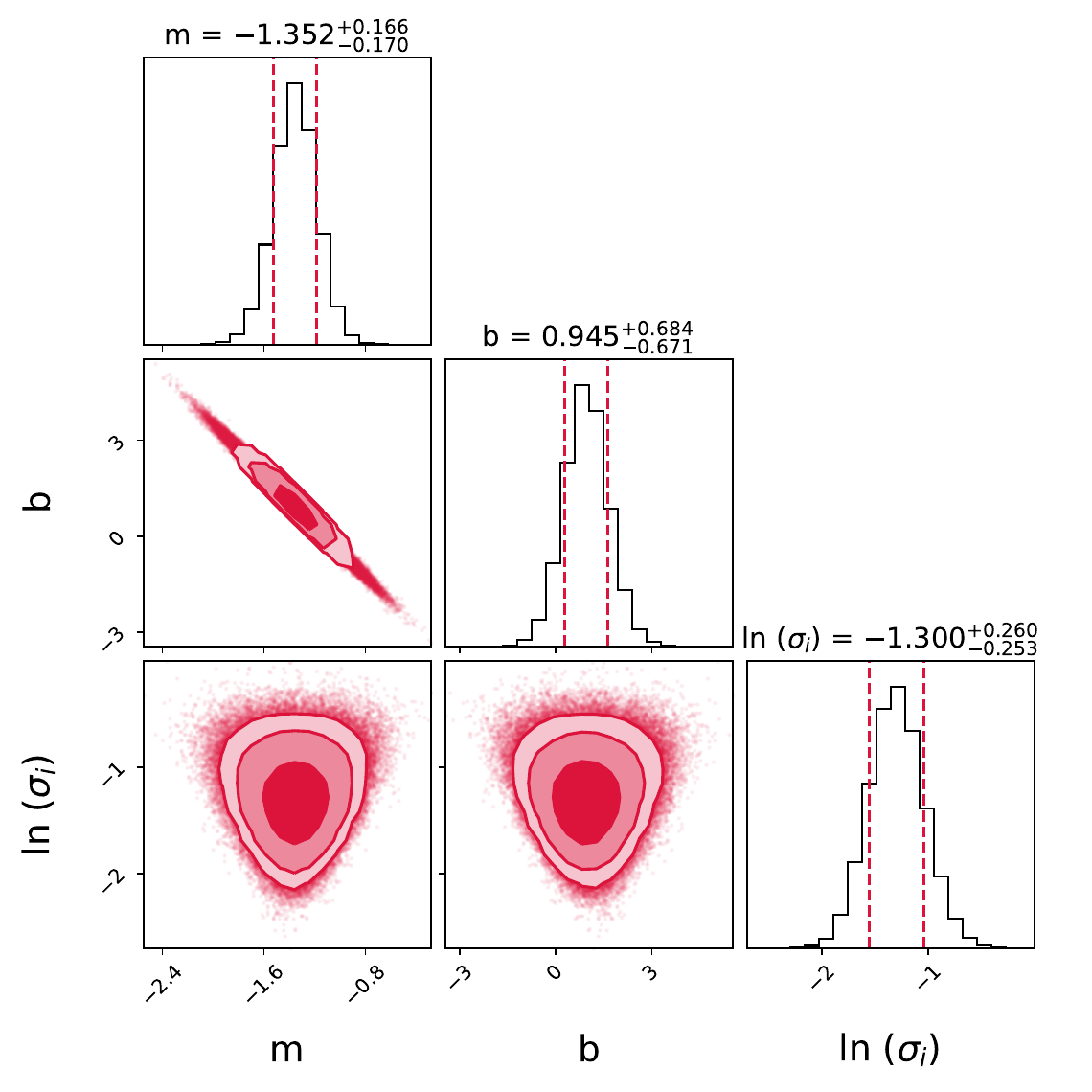}
    \caption{Plot showing the  68\%,  90\% and 99\% marginalized credible intervals  for $m$ (slope), $b$ (intercept), and $\ln \sigma_i$(intrinsic scatter) for the linear relation between $\rho_{c}$ and $r_{c}$ in log-log space.}
    \label{fig:f2}
\end{figure*}

The maximization of the likelihood is done using the {\tt emcee} MCMC sampler~\cite{emcee}.
The best-fit value for the scaling relation between $\rho_c$ and $r_c$, which we obtained is given by:

\begin{eqnarray}
\ln\bigg(\frac{\rho_c}{M_{\odot} pc^{-3}}\bigg)=(-1.35^{+0.16}_{-0.17}) \ln\bigg(\frac{r_c}{kpc}\bigg)\nonumber\\
+(0.94^{+0.66}_{-0.68})  
\label{eq:eq9} 
\end{eqnarray}
with an intrinsic scatter of $(27.3^{+9.2}_{-8.9})\%$. The best fit values and their $1\sigma$ levels for the slope ($m$), intercept ($b$)  are superposed in Figure \ref{fig:f2} along with the data. We find that the intrinsic scatter for this scaling relation is about a factor of 1.5 times larger than what was seen for the  Chandra cluster sample analyzed in GD20. Therefore, the relation between $\rho_c$ and $r_c$ is not as tight as seen for galaxy clusters.  We find a deviation from a constant dark matter halo density at about 2.0$\sigma$.  We also redid the analysis by omitting the RGH 80 group data (given that it may not be in hydrostatic equilibrium),  and obtained a slope of $-1.40^{+0.18}_{-0.17}$ with an intrinsic scatter of 27.3\%.  Therefore, we find that this group does not make an appreciable difference to the final results.

\rthis{Finally, the median value of $S_{surf}$  for our sample is equal to  612 $M_{\odot}/pc^2$, which is about four times larger than the values for single galaxies~\cite{Salucci19}, but still smaller  than what was obtained for galaxy clusters in GD20. Therefore, this shows that the halo surface density cannot be a constant across all scales.}

\begin{table}[h]
\begin{ruledtabular}
\begin{tabular}{ccc}
Galaxy Group & $\rho_c$ & $r_c$ \\
& $10^{-3}M_{\odot} pc^{-3}$ & kpc \\\hline

IC 1860  & 9.33 $\pm 0.49 $ &  53.34 $\pm 1.65 $ \\ 
A262  & 15.49 $\pm 1.30 $ &  53.38 $\pm 2.19 $ \\ 
A2717  & 4.36 $\pm 0.24 $ &  115.92 $\pm 5.02 $ \\ 
AWM 4  & 9.19 $\pm 1.01 $ &  76.96 $\pm 6.16 $ \\ 
ESO 3060170  & 17.70 $\pm 0.56 $ &  56.10 $\pm 2.22 $ \\ 
ESO 5520200  & 3.21 $\pm 0.52 $ &  120.87 $\pm 12.54 $ \\ MKW 4  & 21.76 $\pm 3.49 $ &  41.73 $\pm 2.70 $ \\ 
MS 0116.3-0115  & 7.09 $\pm 0.93 $ &  71.09 $\pm 6.29 $ \\ 
NGC 533  & 12.37 $\pm 3.50 $ &  39.61 $\pm 4.29 $ \\ 
NGC 1550  & 45.72 $\pm 13.10 $ &  21.86 $\pm 2.78 $ \\ 
NGC 2563  & 11.49 $\pm 5.07 $ &  38.11 $\pm 6.88 $ \\ 
NGC 4325  & 17.76 $\pm 2.54 $ &  35.15 $\pm 3.36 $ \\ 
NGC 5129  & 8.13 $\pm 0.99 $ &  42.63 $\pm 4.71 $ \\ 
RGH 80  & 21.68 $\pm 3.57 $ &  28.24 $\pm 1.94 $ \\ 
RX J1159.8+5531  & 18.75 $\pm 3.01 $ &  47.03 $\pm 4.07 $ \\ 
NGC 5044  & 87.32 $\pm 27.73 $ &  16.97 $\pm 2.36 $ \\ 
A2589  & 6.29 $\pm 0.15 $ &  83.11 $\pm 1.15 $ \\ 
\end{tabular}
\end{ruledtabular}
\caption{\label{tab:table1} Estimated values for the core density ($\rho_c$) and the core radius ($r_c$)  for the Z06 and  G07 galaxy group sample. \rthis{The median value of their product,  $S_{surf}$ is equal to 612 $M_{\odot}/pc^2$.}}
\end{table}

\section{Radial Acceleration Relation in Groups}
\label{sec:RAR}
\begin{figure}
    \includegraphics[scale=0.50]{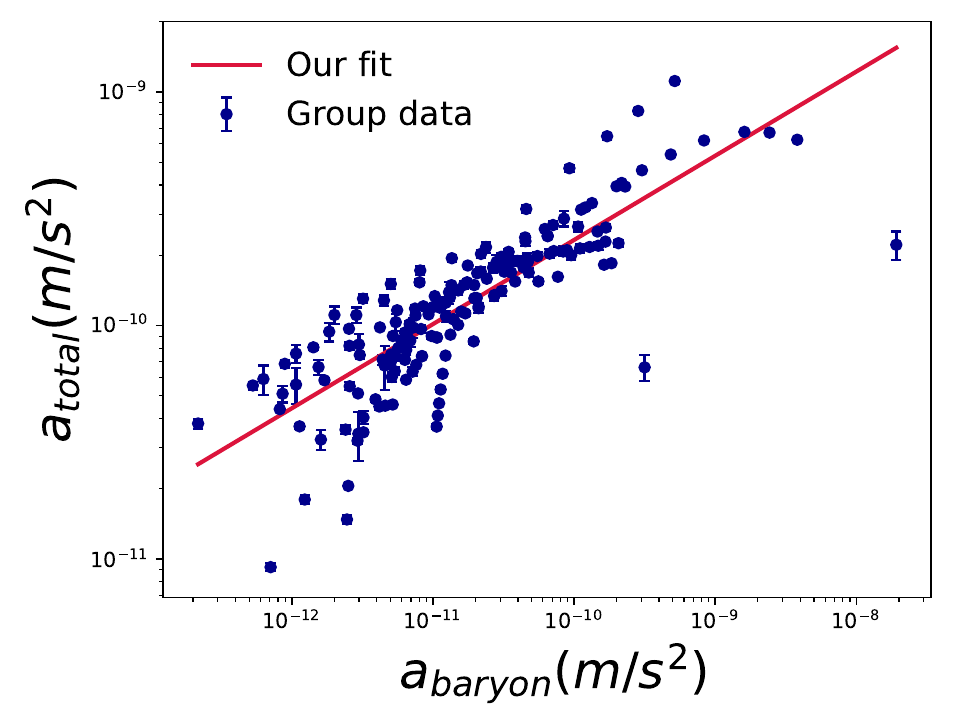}
    \caption{A test of  the RAR using  the 17 groups from Z06 and G07 sample. The crimson solid line is the fit from our analysis. }
    \label{fig:f3}
\end{figure}

We have also used this same dataset  to  implement a test for the RAR, similar to the analysis in ~\cite{Pradyumna} (P21, hereafter). The estimated total and baryonic acceleration for  the radii for which  the data and errors  were available to us was stacked together. A plot showing the relation between the two  in the logarithmic space can be found  in Figure~\ref{fig:f3}.

To test the RAR, a linear regression was done by maximizing the log-likelihood function (Eq.~\ref{eq:eq8}) in the same way as was done in the previous sub-section. Here, $x \equiv \ln(a_{tot})$ and $y \equiv \ln (a_{baryon})$, and their errors are denoted  by  $\sigma_x$ and $\sigma_y$, respectively as before. We again incorporate an intrinsic scatter  as a free parameter. We obtained $a_{tot}\propto a_{baryon}^{0.36\pm0.02}$ with an intrinsic scatter of 42\%. The best-fit results are tabulated in Table~\ref{tab:summary}, and can be compared with the results from other cluster data. The 68\%, 90\%, and 99\% marginalized confidence intervals for $m$, $b$, and $\sigma_{intrinsic}$ can be found in Fig. 4. 

\begin{figure*}[]
    %\centering
    \includegraphics[scale=0.85]{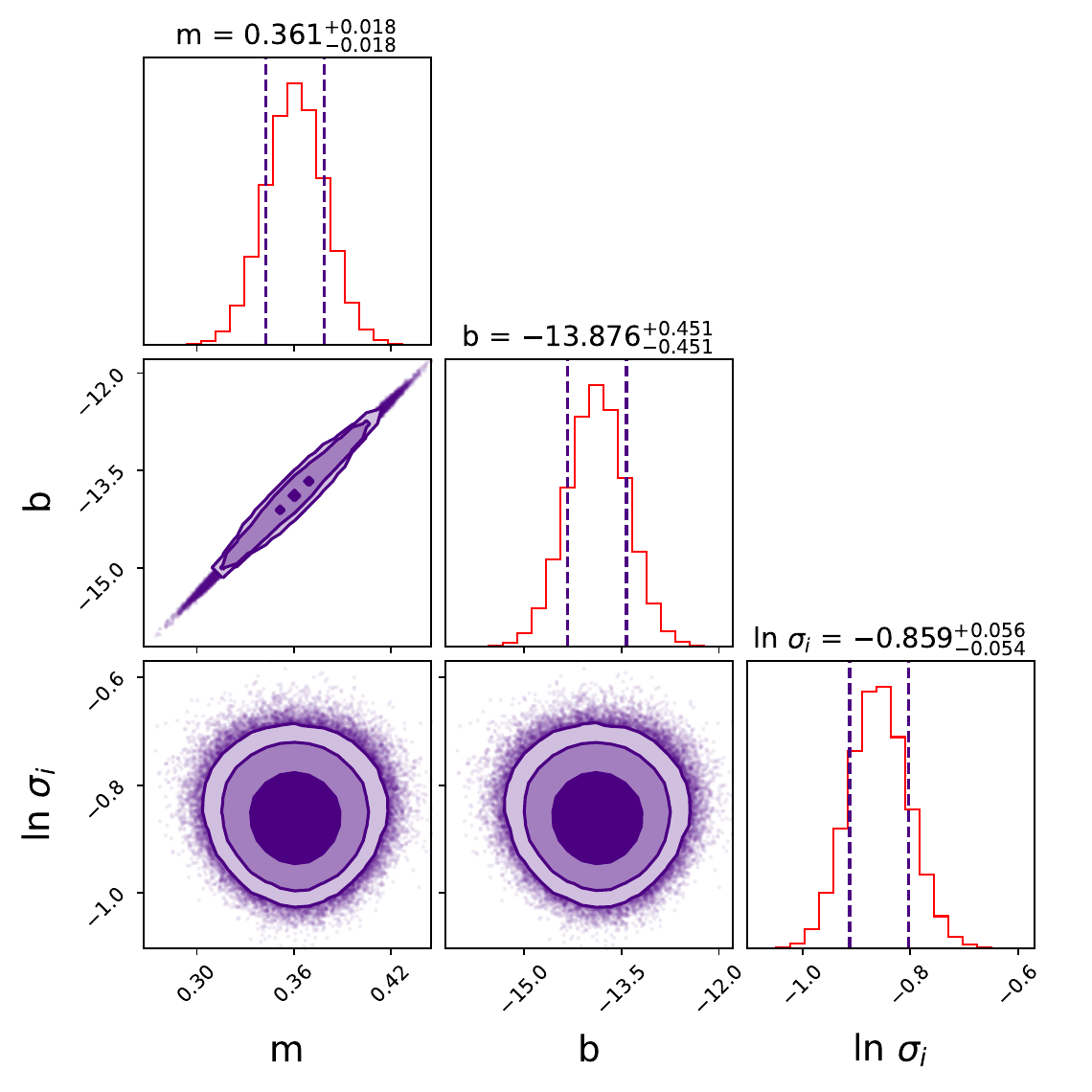}
    \caption{Plot showing the  68\%,  90\% and 99\% marginalized credible intervals  for $m$ (slope), $b$ (intercept), and $\ln  \sigma_i$(intrinsic scatter) for the linear relation between $a_{tot}$ and $a_{baryon}$ in log-log space.}
    \label{fig:f4}
\end{figure*}

\begin{table*}[]
\begin{ruledtabular}
\begin{tabular}{lccccc}
Slope & Intercept & Intrinsic Scatter & Residual Scatter(dex) & $a_0(m/s^2)$ & Data Sample \\ \hline

$0.36^{+0.02}_{-0.02}$ & $-13.87^{+0.45}_{-0.45}$ & $0.42\pm{0.02}$ & 0.32 & $(0.88\pm0.06)\times 10^{-9}$ & XMM and Chandra (this work) \\

$1.09^{+0.07}_{-0.07}$ & $4.21^{+0.1}_{-0.1}$ & $1.36\pm0.02$ & 0.11 & $(1.12\pm0.11)\times 10^{-9}$ & XCOP\cite{Pradyumna}\\

$0.51^{+0.04}_{-0.05}$ & $-10.17^{+0.03}_{-0.03}$ & $0.49 \pm0.02$ & 0.11 & $(2.02\pm0.11)\times 10^{-9}$ & CLASH\cite{Tian} \\
\end{tabular}
\end{ruledtabular}
\caption{\label{tab:summary}Summary of results for a linear regression of  $\ln(a_{tot})$ versus $\ln (a_{baryon})$ from different group and cluster samples.}
\end{table*}

% \begin{eqnarray}
% \ln\bigg(\frac{a_{tot}}{m/s^2}\bigg)=(0.35^{+0.02}_{-0.02}) \ln\bigg(\frac{a_{baryon}}{m/s^2}\bigg)\nonumber\\
% -(14.10^{+0.50}_{-0.51})  
% \label{eq:eq10} 
% \end{eqnarray}

The residual scatter in the RAR for this group sample is $\sigma=0.32$ dex. Similar to P21, this was obtained by fitting a Gaussian function to the frequency distribution of the difference between $\log_{10} (a_{obs})$ and $\log_{10} (a_{expected})$. The standard deviation of the fitted Gaussian function gives the residual scatter in dex, which is used to characterize the tightness of RAR relation of this sample. As we can see, the residual scatter  obtained (0.32 dex) is about three times  higher than what was obtained with the cluster samples~\cite{Pradyumna,Tian,Pradyumna2} or for galaxies~\cite{Li18}. Therefore, we conclude that our sample of galaxy groups does not show the same tight scatter  seen previously for other astrophysical systems.

\par The acceleration scale ($a_0$) is evaluated in the same way  as in P20 (See also ~\cite{Tian}). The slope for the linear regression is fixed to  0.5 in the regression relation, thereby recasting  the RAR relation  as $a_{dyn}=\sqrt{a_0 a_{baryon}}$. The best-fit  value with this assumption for $a_0$ is $(0.88\pm{0.06})\times10^{-9}m/s^2$, which is slightly lower than what was obtained for CLASH~\cite{Tian} and XCOP~\cite{Pradyumna} data, but still larger than what was seen for the SPARC sample~\cite{Mcgaugh16}.

\section{Correlation with mass and Luminosity}
\label{sec:corr}
We now use our estimates for $S_{surf}$ to test for any putative correlation with mass and luminosity, which have been reported using observational data on galactic scales~\cite{DelPopolo12,Kormendy14,DelPopolo20}, and also  to test the predictions of some  theoretical models~\cite{DelPopolo12,Loeb}.
We carry out a regression analysis in log-log space between each of the observables. A summary of these results  along with the intrinsic scatter for each of these relations is summarized in Table \ref{tab:M&L summary}.

\begin{figure}
    \includegraphics[scale=0.50]{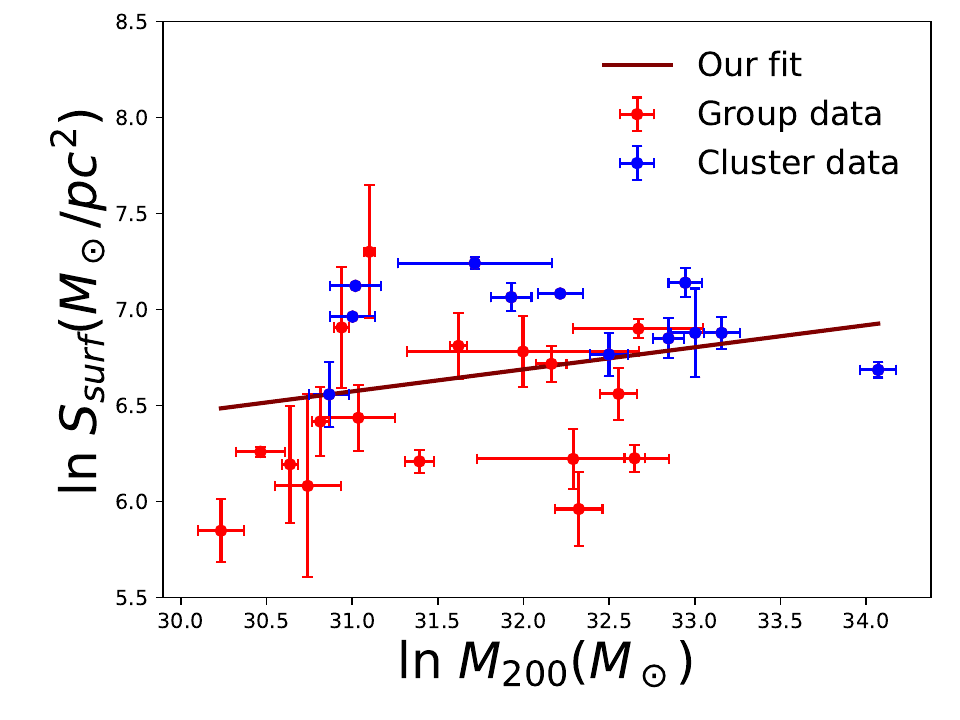}
    \caption{Correlation of dark matter halo surface density with mass using data for galaxy groups analyzed in this work and galaxy clusters from GD20. The black solid line shows the fit from our analysis. The best-fit values can be found in Table~\ref{tab:M&L summary}. }
    \label{fig:f5}
\end{figure}

 To check for a correlation of $S_{surf}$ with the halo mass,  similar to GD20, we use $M_{200}$ as a proxy, where $M_{200}$ is the mass at the over-density $\Delta=200$ with respect to the critical density. $M_{200}$ values were available in G07 and Z06, which were obtained through NFW~\cite{NFW} fits.
The dark matter surface density ($S$) for each of the groups as a function of $M_{200}$  is shown in Figure \ref{fig:f5}.  While testing for a correlation between the two, we incorporated the cluster data from GD20 in order to increase the dynamical range for the mass.  A marginal positive correlation ($S\propto M_{200}^{0.11\pm0.07}$), can be seen  between these  observables, which is in accord  with some of the previous results  seen at galactic scales~\cite{DelPopolo12,DelPopolo20}. However, additional data is necessary to confirm if this correlation is significant.
From Figure~\ref{fig:f5}, we also find that the dark matter surface density of groups can be delineated from that of   clusters and falls in the left corner of the plot.
%Our best-fit relation are tabu 
% \begin{eqnarray}
% \ln\bigg(\frac{S}{M_\odot/pc^2}\bigg)=(0.11\pm 0.07) \ln\bigg(\frac{M_{200}}{M_\odot}\bigg)\nonumber\\
% +(3.01\pm2.4)  
% \label{eq:eq10} 
% \end{eqnarray}
% with an intrinsic scatter of $35.6^{+5.5}_{-5.9}\%$.

\begin{figure}
    \includegraphics[scale=0.50]{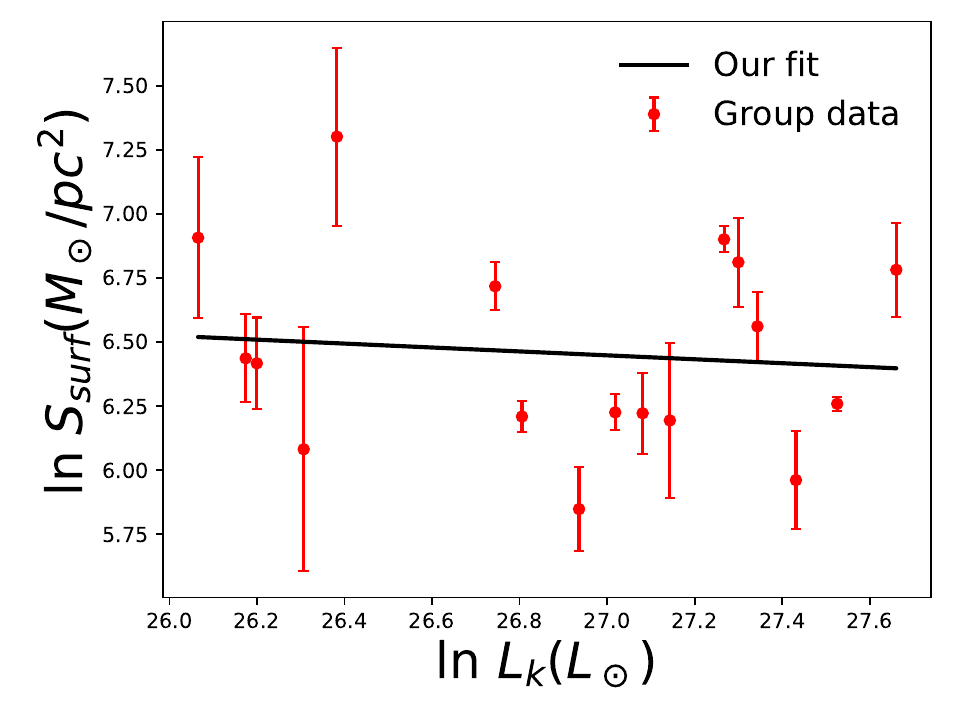}
    \caption{Correlation of halo surface density with $K$-band luminosity. The best-fit values can be found in Table~\ref{tab:M&L summary}.}
    \label{fig:f7}
\end{figure}

Previously, contradictory results have been obtained for the correlation between halo surface density and luminosity (or  magnitude). The analysis in ~\cite{Donato,Kormendy14} found no correlation between the surface density and $B$-band absolute magnitude. However, most recently, ~\citeauthor{DelPopolo20} (see also ~\cite{Saburova}) showed that for the SPARC sample, the surface density is correlated  with  the luminosity measured at 3.6$\mu$m.
Our data for $S$ as a function of $L_K$ is shown in Fig.~\ref{fig:f7}. We find that $S\propto L^{-0.08\pm0.19}_k$. Therefore, although we find that there is no evolution of $S$ with $L_k$ within 1$\sigma$, we also cannot rule out any correlation (or anti-correlation) with the luminosity. Additional data for group scale haloes is  necessary to make a definitive statement.

\begin{figure}
    \includegraphics[scale=0.50]{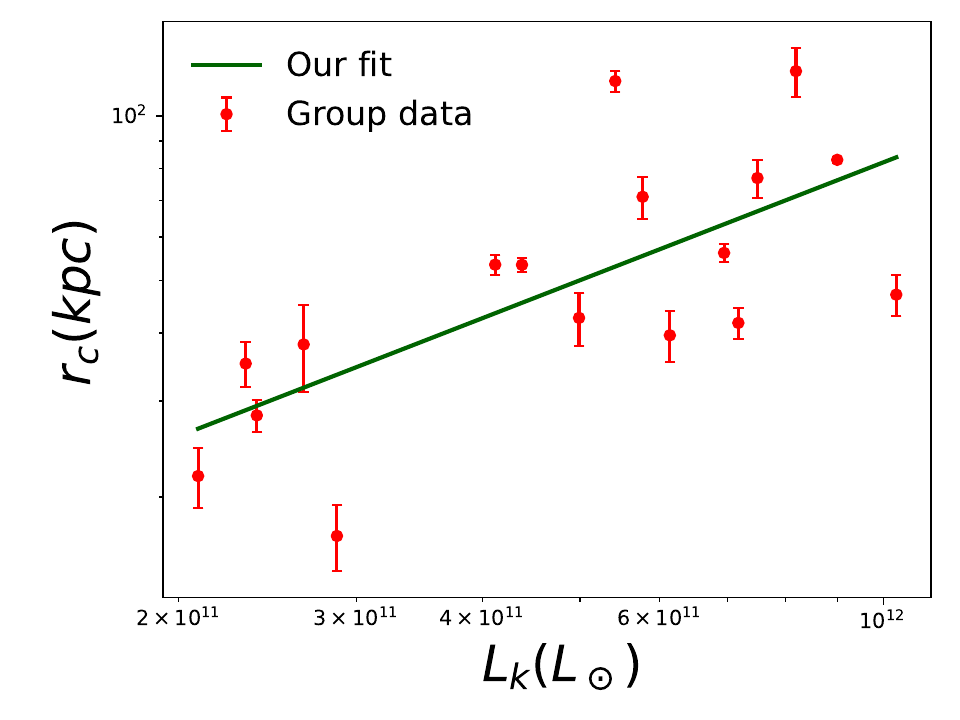}
    \caption{Correlation of $K$-band luminosity with the halo core radius. The best-fit values can be found in Table~\ref{tab:M&L summary}.}
    \label{fig:f6}
\end{figure}

Previously, a correlation between  $r_c$ and the   stellar mass was found, using the   $B$-band luminosity  ($L_B$)  as  a proxy~\cite{Kormendy14}.   Their analysis showed that $r_c \propto L_{B}^{0.446}$. Here,  since we are using $L_K$
as a proxy for stellar mass, we check for correlation between $r_c$ and $L_K$. This data is shown in Fig.~\ref{fig:f6}. There is a slight positive correlation between these quantities, $r_c \propto L_K^{0.71 \pm 0.2}$, in agreement with these results.

We should also caution that the intrinsic scatter for all the aforementioned relations (cf. Table~\ref{tab:M&L summary}) is also quite high $>30\%$.  Therefore, for a more definitive test, much more data would be needed to ascertain some of these correlations and check if the scatter becomes tighter with more statistics.

% \begin{eqnarray}
% \ln\bigg(\frac{r_c}{kpc}\bigg)=(0.71^{+0.20}_{-0.19}) \ln\bigg(\frac{L_{k}}{L_\odot}\bigg)\nonumber\\
% -(15.396^{+5.2}_{-5.4})  
% \label{eq:eq11} 
% \end{eqnarray}
% where, there is an intrinsic scatter of 39\%.

\begin{table*}[]
\begin{ruledtabular}
\begin{tabular}{lccc}
Slope & Intercept & Intrinsic Scatter(\%) & Scaling Relation \\ \hline

$0.11^{+0.07}_{-0.07}$ & $3.0^{+2.43}_{-2.42}$ & $35.5^{+5.5}_{-5.9}$ & $S_{surf}-M_{200}$ \\

$-0.08^{+0.19}_{-0.19}$ & $8.61^{+5.14}_{-5.30}$ & $32.5^{+8.6}_{-9.0}$ & $S_{surf}-L_{k}$ \\

$0.71^{+0.20}_{-0.19}$ & $-15.39^{+5.2}_{-5.4}$ & $39.1^{+7.5}_{-6.4}$ & $r_c-L_{k}$ \\

$0.13^{+0.09}_{-0.09}$ & $1.39^{+2.92}_{-3.14}$ & 
$24.8^{+8.1}_{-7.7}$ & $S_{Bur}-M_{200}$ \\
\end{tabular}
\end{ruledtabular}
\caption{\label{tab:M&L summary} Summary of the study of possibly correlation of the halo surface density ($S$) and core radius ($r_c$) with mass and luminosity. In each of these cases we fit the observables using linear regression in log-log space}
\end{table*}

\section{Correlation of dark column density with mass}
\label{sec:sburk}
\rthis{We now calculate a variant of the dark matter surface density called dark matter ``column density'',  for a dark matter density profile ($\rho_{DM}$) defined at  radius ($R$)~\cite{Boyarsky,BoyarskyPRL}}
\begin{equation}
S (R)  = \frac{2}{R^2} \int_0^{R} r' dr' \int_{-\infty}^{+\infty} dz \rho_{DM} (\sqrt{r'^2+z^2})
\label{eq:SB}
\end{equation}
\rthis{This measure of surface density is more robust  than $S_{surf}$, as it has been shown to be invariant with respect to the underlying dark matter density profile used~\cite{Boyarsky,DelPopolo12}. Therefore, this column density has also been used as an alternative  probe of average dark matter surface density in literature. 
The    column densities for  NFW ($S_{NFW}$), pseudo-isothermal ($S_{ISO}$) and Burkert profiles ($S_{BUR}$) are linked with each other according to : $S_{NFW} (r_s) \approx  0.91 S_{ISO} (6 r_c) $ and $S_{NFW} (r_s)  \approx 0.98  S_{BUR} (1.66 r_0)$~\cite{Boyarsky}. The analytical expressions for  $S_{NFW}$ and $S_{ISO}$  have been provided in ~\cite{Boyarsky,DelPopolo12}.}

\begin{figure}
    \includegraphics[scale=0.50]{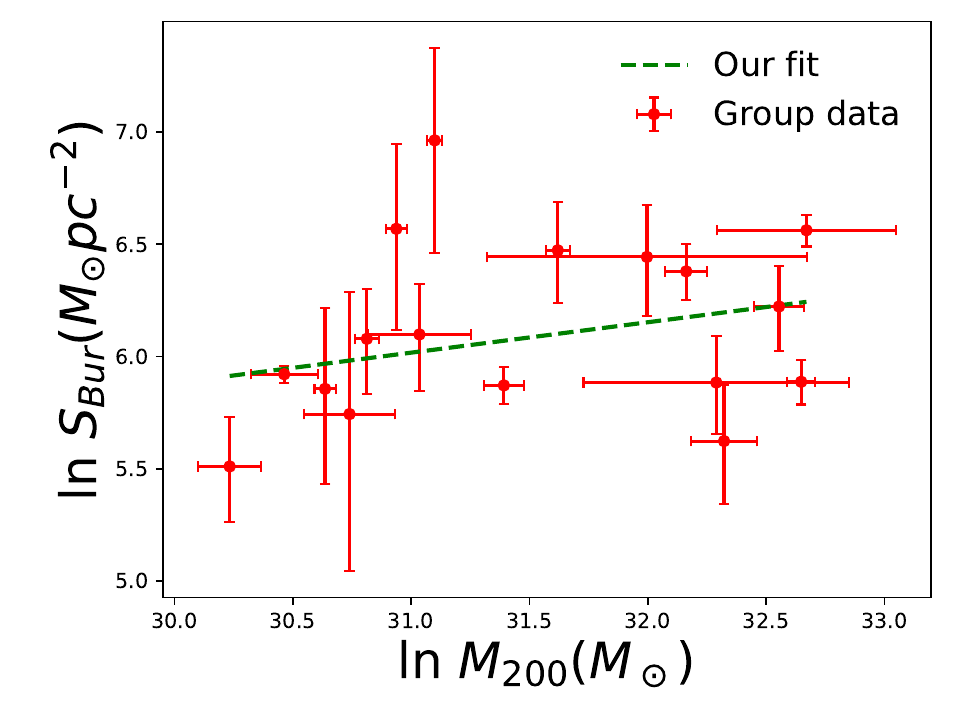}
    \caption{\rthis{Correlation of the dark matter column density calculated for the Burkert profile (cf. Eq.~\ref{eq:SB}) with halo mass. The best-fit values can be found in Table~\ref{tab:M&L summary}.}}
    \label{fig:f8}
\end{figure}

\rthis{The values of $S_{NFW}$ for all the galaxy groups analyzed in this work have been tabulated in ~\cite{Boyarsky}, who find that $S_{NFW} \propto M_{halo}^{0.2}$. Here, we calculate $S_{BUR}$  using the $\rho_c$ and $r_c$ data obtained for each group from Table ~\ref{tab:table1}. $S_{BUR}$ was calculated by plugging in the Burkert density profile in Eq.~\ref{eq:SB}, and numerically evaluating the integrals. The values of $S_{BUR}$ for all the groups of our sample are  plotted as a function of $M_{200}$ in Fig.~\ref{fig:f8}. $S_{BUR}$ is of the same order of magnitude as $S_{surf}$, with their median ratio equal to 0.71.
We carry out a regression analysis in the same way as in Sect.~\ref{sec:corr}. Our best-fit values are tabulated in Table~\ref{tab:M&L summary}. Similar to $S_{surf}$, we find a mild correlation with halo mass ($S_{burk} \propto M_{200}^{0.13 \pm 0.09}$). However,  the intrinsic scatter is about 25\%, which although  is slightly smaller than that obtained for the halo surface density is still quite large.}

\section{Conclusions}
\label{sec:conclusions}
In this work, we implemented  a test for  the  constancy of the  dark matter halo surface density \rthis{as well as column density}   and  RAR  for a sample of 17 galaxy groups observed using Chandra and XMM-Newton.

Several studies done previously with galaxies unveiled a very intriguing  result that the dark matter halo surface density, given by the product of $\rho_c$ and $r_c$, obtained from fitting a cored profile, is a constant irrespective of the mass, spectral type, and blue magnitude~\cite{Donato,Kormendy14}. \rthis{However, this result assumes that dark matter density profiles can be well-fitted by cored profiles, and consequently has been disputed by several other works, which find a correlation with halo mass.}
The observed halo surface density has  proved to be an invaluable probe of  alternatives to the standard $\Lambda$CDM model~\cite{Chan,Burkert20}. 

Very few studies  of this {\it ansatz} were previously done with galaxy clusters, apart from   a detailed analysis   of a  ROSAT cluster sample~\cite{Chan}, which showed that galaxy clusters do not have a constant surface density. Recently, GD20  carried out  a systematic  test  using  a sample of  12 Chandra clusters, and found that $\rho_c\propto r_c^{-1.08^{+0.05}_{-0.06}}$, indicating a deviation from a constant surface density by about 1.4$\sigma$. \rthis{They also showed that the halo surface density for the cluster sample is about ten times larger than that for galaxies.}

Here, we have extended the same analysis  to lower mass systems, using  a set of 17 relaxed galaxy groups from XMM-Newton and Chandra X-ray observations. We find that the dark matter core density ($\rho_c$) and core radius ($r_c$) obey the   scaling relation, $\rho_c\propto r_c^{-1.35 \pm 0.17}$.
However, the intrinsic scatter of 27\% which we obtain is about 1.5 times larger than that seen for clusters, indicating that the relation is not as tight as seen for galaxy clusters. Furthermore, the halo surface density is discrepant with respect to a constant value by about 2.0$\sigma$. 
The dark matter surface density which we obtained  for the groups falls just below the clusters \rthis{and is about four times larger than that obtained for galaxies~\cite{Salucci19}.}

 We have also tested for a  correlation of both the  surface density and \rthis{column density} with mass and obtained a marginal positive correlation (albeit with a large scatter), which is in agreement with some previous estimates~\cite{DelPopolo12}.  We cannot confirm or rule out a correlation of the surface density with $K$-band luminosity.
 We also find that the core radius is correlated with $K$-band luminosity, in agreement with the results in ~\cite{Kormendy14}.

As a follow-up to another recent work of ours which tested the RAR for three different cluster samples~\cite{Pradyumna,Pradyumna2}, we  also implemented a test  of RAR using the same group sample to see how well RAR holds up at group scales. Our results are summarized in Table~\ref{tab:summary}. We find that the residual scatter in the RAR relation for groups is about 0.3 dex, and subsequently about three times larger than that seen for galaxies or clusters. Therefore, galaxy groups do not obey the RAR. The acceleration scale which we obtain assuming a slope of 0.5, is between the values seen for clusters and single galaxies.

Additional tests for the halo surface density and RAR with additional group catalogs (for eg~\cite{Sun09}) and the upcoming e-ROSITA sample, followed by comparison with predictions from $\Lambda$CDM and alternatives will be discussed in future works. 

\rthis{Nevertheless, the result in this work in conjunction with G20, prove  that the dark matter halo surface density cannot be a  universal constant at all scales or for all dark matter dominated systems.}

\section*{Acknowledgements}

We are grateful to Fabio Gastaldello for useful correspondence  and for providing us the data in G07. We would also like to thank Man-Ho Chan, Antonio Del Popolo,  Yong Tian, and S. Pradyumna  for helpful discussions on this topic.  \rthis{Finally, we thank the anonymous referee for useful comments and feedback on our manuscript.}

\clearpage

\bibliography{new}
\end{document}